\documentclass[twocolumn,epjc3]{svjour3}

\RequirePackage{graphicx}

\RequirePackage{amsmath,amsfonts,amssymb}
\RequirePackage[numbers,sort&compress]{natbib}
\RequirePackage[colorlinks,citecolor=blue,urlcolor=blue,linkcolor=blue]{hyperref}
\RequirePackage{dsfont}

\usepackage[utf8]{inputenc}

\journalname{Eur. Phys. J. C}

\begin{document}

\title{Cosmological implications of the transition from the false vacuum to the true vacuum state}

\author{Aleksander Stachowski\thanksref{oauj,e-as}
\and
Marek Szyd{\l}owski\thanksref{oauj,csrc,e-ms}
\and
Krzysztof Urbanowski\thanksref{ipuzg,e-ku}}

\thankstext{e-as}{aleksander.stachowski@uj.edu.pl}
\thankstext{e-ms}{marek.szydlowski@uj.edu.pl}
\thankstext{e-ku}{K.Urbanowski@if.uz.zgora.pl}

\institute{Astronomical Observatory, Jagiellonian University, Orla 171, 30-244 Krakow, Poland \label{oauj}
\and
Mark Kac Complex Systems Research Centre, Jagiellonian University, {\L}ojasiewicza 11, 30-348 Krak{\'o}w, Poland \label{csrc}
\and
Institute of Physics, University of Zielona G{\'o}ra, Prof. Z. Szafrana 4a, 65-516 Zielona G{\'o}ra, Poland \label{ipuzg}}

\date{Received: date / Accepted: date}

\maketitle

\begin{abstract}
We study the cosmology with the running dark energy. The parametrization of
dark energy with the respect to the redshift is derived from the first
principles of quantum mechanics. Energy density of dark energy is obtained from
the quantum process of transition from the false vacuum state to the true
vacuum state. This is the class of the extended interacting $\Lambda$CDM
models. We consider the energy density of dark energy parametrization
$\rho_\text{de}(t)$, which follows from the Breit-Wigner energy distribution
function which is used to model the quantum unstable systems. The idea that
properties of the process of the quantum mechanical decay of unstable states
can help to understand the properties of the observed universe was formulated
by Krauss and Dent and this idea was used in our considerations. In the
cosmological model with the mentioned  parametrization there is an energy
transfer between the dark matter and dark energy. In such a evolutional
scenario the universe is starting from the false vacuum state and going to
the true vacuum state of the present day universe. We find that the intermediate
regime during the passage from false to true vacuum states takes place. 
The intensity of the analyzed process is measured by a parameter $\alpha$.
For the small value of $\alpha$ ($0<\alpha <0.4$) this intermediate (quantum) regime is
characterized by an oscillatory behavior of the density of dark energy
while the for $\alpha > 0.4$ the density of the dark energy simply jumps
down. In both cases (independent from the parameter $\alpha$) the today value
of density of dark energy is reached at the value of $0.7$. We estimate the cosmological
parameters for this model with visible and dark matter. This model becomes in
good agreement with the astronomical data and is practically indistinguishable
from $\Lambda$CDM model.
\end{abstract}

\section{Introduction}
The standard cosmological model ($\Lambda$CDM model), which describes the Universe, is the most favored by astronomical observations such as supernovae of type Ia or measurements of CMB. In the $\Lambda$CDM model, the dark matter is treated as dust and dark energy has the form of the cosmological constant $\Lambda_\text{bare}$. We are looking an alternative for the $\Lambda$CDM model by a modification of the dark energy term.

The standard cosmological model possesses the six parameters:
the density of baryons $\Omega_\text{b}h^2$,
the density of cold dark matter $\Omega_\text{dm} h^2$,
angular diameter of sound horizon at last scattering $\theta$,
the optical depth due to the reionisation $\tau_\text{R}$,
the slope of the primordial power spectrum of fluctuations $n_\text{s}$, and
the amplitude of the primordial power spectrum $A_\text{s}$, where $h = H_0$ [100 km s${}^{-1}$ Mpc${}^{-1}$].

From the methodological point of view, the standard cosmological model plays the role of an effective theory, which very well describes properties of the current Universe without explaining the nature of two components of the model: the dark energy and the dark matter. The nature of both components of the Universe is unknown up to now but we describe these in terms of useful fiction, the cosmological constant and the cold dark matter which is kind of the dust perfect fluid.

In this paper we concentrate on the interpretation of dark energy rather in terms of running cosmological constant than in term of the pure cosmological constant parameter ($\Lambda_{\text{bare}}$ in our approach). It is consequence of some problems with interpretation of the pure cosmological constant, namely:
\begin{enumerate}
\item One cannot explain why the cosmological constant is not large.
\item One do not know why it is not just equal zero.
\item One cannot explain why energy densities of both dark energy and dark matter, expressed in terms of dimensionless density parameters, are comparable  in the current epoch (cosmic coincidence problem).
\end{enumerate}

In our proposition of the explanation of these problems with the cosmological constant parameter, we base on the theories of the cosmological constant in which the vacuum energy is fixed by the fundamental theory \cite{Polchinski:2006gy}. Extending the $\Lambda$CDM model beyond the classical regime, we apply quantum mechanics as a fundamental theory, which determines cosmological parameters and explain how cosmological parameters vary during the cosmic evolution.

The cosmological constant is the source of two problems in modern cosmology. The first problem is the cosmological constant problem, which is consequence of the interpretation of dark energy as a vacuum energy. The observed present value of the cosmological constant is 120 orders of magnitude smaller than we expect from quantum physics. The second problem is the coincidence problem. If we assume that the dark energy is always constant then the $\Lambda$CDM model cannot explain why the cosmological constant has the same order of magnitude as density of matter today. If the model belongs to the class of running dark energy cosmologies then the first problem of cosmological constant can be solved.

This question seems to be crucial in contemporary physics because its solution would  certainly mean a very crucial step forward in our attempts to understand physics from the boundary of particle physics and cosmology. The discussion about the cosmological constant problem can be found in papers \cite{Polchinski:2006gy,Weinberg:1988cp,Carroll:1991mt,Dolgov:1997za,Sahni:1999gb,Straumann:1999ia,Weinberg:2000yb,Carroll:2000fy,Rugh:2000ji,Padmanabhan:2002ji, Yokoyama:2003ii,Sarkar:2005bc,Copeland:2006wr,Szydlowski:2015bwa,Bousso:2013uia,Bousso:2014jca}.

In our model, the influence of running dark energy densities of both visible and invisible matter is very small. Thus we share Weinberg's opinion, according to which looking for a solution of the coincidence problem, we should consider the anthropic principle. According to Weinberg's argument, any observers should not be in the Universe if the cosmological constant was even three orders of magnitude larger than it is now.

Coleman {\it et al}.~\cite{Coleman:1977py,Callan:1977pt,Coleman:1980aw} discussed the instability of a physical system, which is not at an absolute energy minimum, and which is separated from the minimum by an effective potential barrier. They showed that if the early Universe is too cold to activate the energy transition to the minimum energy state then a quantum decay, from the false vacuum to the true vacuum, is still possible through a barrier penetration via the macroscopic quantum tunneling.

The discovery of the Higgs-like resonance at 125-126 GeV \cite{Kobakhidze:2013tn,Degrassi:2012ry,EliasMiro:2011aa,Chao:2012mx} caused the discussion about the instability of the false vacuum. If we assume that the Standard Model well describes the evolution of the Universe up to the Planck epoch then a Higgs mass $m_\text{h}< 126 \text{GeV}$ causes that the electroweak vacuum is in a metastable state \cite{Degrassi:2012ry}. In consequence the instability of the Higgs vacuum should be considered in the cosmological models of the early time Universe.

The idea that properties of the quantum mechanical decay process of metastable states
can help to understand
the properties of the observed universe was formulated in \cite{Krauss:2007rx,Krauss:2008pt,Winitzki:2007cf}. It is because
the decay of the false vacuum  is the
 quantum decay process \cite{Coleman:1977py,Callan:1977pt,Coleman:1980aw}.
 This means that  state vector corresponding  to the false vacuum is
a quantum unstable (or metastable) state. Therefore
 all  general properties of quantum unstable systems  must also occur
 in the case of such a quantum unstable state as the false vacuum and as a consequence models of quantum unstable systems can be used to analyze
 properties of the systems which time evolution starts from the false vacuum state.

In this paper, we assume the Breit-Wigner energy distribution function, which is very often used to model unstable quantum systems,
as a model of the process of the energy transition from the false vacuum to the true vacuum. In consequence the parametrization  of the dark energy is given by formula
\begin{equation}
\rho_\text{de}=E_0+E_\text{R}\frac{\alpha}{1-\alpha} \Re\left(\frac{J(t)}{I(t)}\right),\label{param}
\end{equation}
where $\alpha$ and $E_\text{R}$ are model parameters describing the variation from the standard cosmological model. The values of $\alpha$ parameter belong to interval $\langle 0,\text{ }1)$. Note that if the $\alpha$ parameter or $E_\text{R}$ is equal zero than the model is equivalent to the $\Lambda$CDM model.

Let $\Lambda_\text{bare}=E_0-E_\text{R}$ then Eq.~(\ref{param}) can be rewritten in the equivalent form
\begin{equation}
\rho_\text{de}=\Lambda_\text{bare}+E_\text{R}\left[1+\frac{\alpha}{1-\alpha} \Re\left(\frac{J(t)}{I(t)}\right)\right].
\end{equation}
Here the units $8\pi G = c =1$ are used.

The functions $J(t)$ and $I(t)$ are defined by the following expressions
\begin{equation}
J(t)=\int_{-\frac{1-\alpha}{\alpha}}^\infty \frac{\eta}{\eta^2+\frac{1}{4}}e^{-i\eta\tau}d\eta,\label{J1}
\end{equation}
\begin{equation}
I(t)=\int_{-\frac{1-\alpha}{\alpha}}^\infty \frac{1}{\eta^2+\frac{1}{4}}e^{-i\eta\tau}d\eta.\label{I1}
\end{equation}

Integrals $J(t)$ and $I(t)$ can be expressed by the exact solutions of these integrals. Formula $J(t)$ is described by the following expression
\begin{multline}
J(\tau)=\frac{1}{2} e^{-\tau/2} \left(-2 i \pi +e^{\tau} \text{E}_1\left(\left[\frac{1}{2}-\frac{i (1-\alpha )}{\alpha }\right] \tau\right) \right. \\ \left. +\text{E}_1\left(\left[-\frac{1}{2}-\frac{i(1-\alpha
) }{\alpha }\right] \tau\right)\right) \label{J11}
\end{multline}
and $I(t)$ is expressed by
\begin{multline}
I(\tau)=2 \pi  e^{\left.-\tau\right/2} \left(1+\frac{i}{2 \pi }\left(-e^{\tau} \text{E}_1\left(\left[\frac{1}{2}-\frac{i (1-\alpha )}{\alpha }\right] \tau\right) \right. \right. \\ \left. \left. +\text{E}_1\left(\left[-\frac{1}{2}-\frac{i
(1-\alpha )}{\alpha }\right]\tau\right)\right)\right), \label{I11}
\end{multline}
where $\tau=\frac{\alpha(E_0-\Lambda_\text{bare})}{\hbar (1-\alpha)}V_0 t$ and $V_0$ is the volume of the Universe in the Planck epoch. In this paper we assume that $V_0=1$. The function $E_1(z)$ is called the exponential integral and is defined by the formula: $E_1(z)=\int_{z}^\infty \frac{e^{-x}}{x}dx$ (see \cite{olver,abramowitz}).

\section{Preliminaries: unstable states}

As it was mentioned in Sec.~1 we will use the parametrization of the dark energy transition from the false vacuum state to the true vacuum state following from the quantum properties of a such process. This process is a quantum decay process, so we need quantities characterizing  decay processes of quantum unstable systems. The main information about properties of quantum unstable systems is contained in their decay law, that is in their survival probability. So if one knows that the system is in the initial unstable state $|\phi\rangle \in {\cal H}$, (${\cal H}$ is the Hilbert space of states of the considered system), which was prepared at the initial instant $t_{0} = 0$, then one can calculate its survival probability (the decay law), ${\cal P}(t)$, of the unstable state $|\phi\rangle$ decaying in vacuum, which equals
\begin{equation}
{\cal P}(t) = |A(t)|^{2}, \label{P(t)}
\end{equation}
where $A(t)$ is  the probability amplitude of finding the system at the
time $t$ in the rest frame ${\cal O}_{0}$ in the initial unstable state $|\phi\rangle$,
\begin{equation}
A(t) = \langle \phi|\phi (t) \rangle . \label{a(t)}
\end{equation}
and $|\phi (t)\rangle$ is the solution of the Schr\"{o}dinger equation
for the initial condition  $|\phi (0) \rangle = |\phi\rangle$, which has the following form
 \begin{equation}
i \hbar \frac{\partial}{\partial t} |\phi (t) \rangle = H |\phi (t)\rangle.  \label{Schrod}
\end{equation}
 Here $|\phi \rangle, |\phi (t)\rangle \in {\cal H}$, and $H$ is the total self-adjoint Hamiltonian for the system considered. The spectrum of $H$ is assumed to be bounded from below
 $E_{\text{min}} > - \infty$ is the lower bound of the spectrum $\sigma_{c}(H) = [E_{\text{min}}, +\infty) $ of $H$).
Using the basis in ${\cal H}$ build from normalized eigenvectors $|E\rangle,\;\ E\in \sigma_{c}(H)$ of $H$ and expanding $|\phi\rangle$ in terms of these eigenvectors
one can express the amplitude $A(t)$ as the following Fourier integral
\begin{equation}
A(t)  \equiv \int_{E_{min}}^{\infty} \omega(E)\;
e^{\textstyle{-\,\frac{i}{\hbar}\,E\,t}}\,d{E},
\label{a-spec}
\end{equation}
where $\omega(E)  > 0$ (see: \cite{fock,Fock:1978fqm,Fonda:1978dk}).

So the amplitude $A(t)$, and thus the decay law ${\cal P}(t)$ of the unstable state $|\phi\rangle$, are completely determined by the density of the energy distribution $\omega(E)$ for the system in this state \cite{fock,Fock:1978fqm}  (see also: \cite{khalfin,Fonda:1978dk,Kelkar:2010qn,muga,muga-1,calderon-2,Giraldi:2015a}. (This approach is also applicable in Quantum Field Theory models \cite{Giacosa:2011xa,goldberger}).

Note that in fact the amplitude $A(t)$ contains information about the decay law ${\cal P}_{\phi}(t)$ of the state $|\phi\rangle$, that is about the decay rate $\Gamma_{\phi}^{0}$ of this state, as well as the energy $E_{\phi}^{0}$ of the system in this state. This information can be extracted from $A(t)$. It can be done using the rigorous equation governing the time evolution in the subspace of unstable states, ${\cal H}_{\parallel} \ni |\phi\rangle_{\parallel} \equiv |\phi \rangle$. Such an equation follows from Schr\"{o}dinger equation (\ref{Schrod}) for the total state space ${\cal H}$.

The use of the Schr\"{o}dinger equation (\ref{Schrod}) allows one to find that within the problem considered
 \begin{equation}
i \hbar \frac{\partial}{\partial t}\langle\phi |\phi (t) \rangle = \langle \phi|H |\phi (t)\rangle.  \label{h||1}
\end{equation}
This relation leads to the conclusion that the amplitude $A(t)$ satisfies the following equation
\begin{equation}
i \hbar \frac{\partial A(t)}{\partial t} = h(t)\,A(t), \label{h||2}
\end{equation}
where
\begin{equation}
h(t) = \frac{\langle \phi|H |\phi (t)\rangle}{A(t)}, \label{h(t)-eq}
\end{equation}
and $h(t)$ is the effective Hamiltonian governing the time evolution in the subspace of unstable states ${\cal H}_{\parallel}= \mathbb{P} {\cal H}$, where
$\mathbb{P} = |\phi\rangle \langle \phi|$ (see \cite{ku-pra} and also \cite{Urbanowski:2006mw,ku-2009} and references therein).
 The subspace ${\cal H} \ominus {\cal H}_{\parallel} = {\cal H}_{\perp} \equiv \mathbb{Q} {\cal H}$ is the subspace of decay products. Here $\mathbb{Q} = \mathbb{I} - \mathbb{P}$. There is the following equivalent formula for $h(t)$ \cite{ku-pra,Urbanowski:2006mw,ku-2009}
\begin{equation}
h(t) \equiv \frac{i\hbar}{A(t)}\,\frac{\partial A(t)}{\partial t}. \label{h(t)}
\end{equation}

One meets the effective Hamiltonian $h(t)$ when one starts with the Schr\"{o}dinger equation for the total state space ${\cal H}$ and looks for the rigorous evolution equation for a distinguished subspace of states ${\cal H}_{||} \subset {\cal H}$ \cite{ku-pra,Giraldi:2015a}. In general $h(t)$ is a complex function of time and in the case of ${\cal H}_{\parallel}$ of dimension two or more the effective Hamiltonian governing the time evolution in such a subspace it is a non-hermitian matrix $H_{\parallel}$ or non-hermitian operator. There is
\begin{equation}
h(t) = E_{\phi}(t) - \frac{i}{2} {\it\Gamma}_{\phi}(t), \label{h-m+g}
\end{equation}
 and
 \begin{equation}
 E_{\phi}(t) = \Re\,[h(t)], \qquad
{\it\Gamma}_{\phi}(t) = -2\,\Im\,[h(t)],\label{m(t)}
\end{equation}
are the instantaneous  energy (mass)  $E_{\phi}(t)$ and the instantaneous decay rate, ${\it\Gamma}_{\phi}(t)$ \cite{ku-pra,Urbanowski:2006mw,ku-2009}. Here $\Re\,(z)$ and $\Im\,(z)$ denote the real and imaginary parts of $z$ respectively. The relations (\ref{h||2}),  (\ref{h(t)}) and (\ref{m(t)}) are convenient when the density $\omega (E)$ is given and one wants to find the instantaneous energy $E_{\phi}(t)$ and decay rate ${\it\Gamma}_{\phi}(t)$: Inserting  $\omega (E)$ into (\ref{a-spec}) one obtains the amplitude $A(t)$ and then using (\ref{h(t)}) one finds the $h(t)$ and thus $E_{\phi}(t)$ and ${\it\Gamma}_{\phi}(t)$. The simplest choice is to take $\omega(E)$ having the Breit-Wigner form
\begin{equation}
\omega (E) \equiv \omega_{\text{BW}}(E) \stackrel{\text{def}}{=} \frac{N}{2\pi}\,
\frac{{\it\Gamma}_{0} {\it\Theta} (E - E_{\text{min}})}{(E-E_{0})^{2} +
(\frac{{\it\Gamma}_{0}}{2})^{2}}, \label{omega-BW}
\end{equation}
where $N$ is a normalization constant and ${\it\Theta} (E) = 1$ for $E\geq 0$ and ${\it\Theta} (E)  =  0$ for $E< 0$. The parameters $E_{0}$ and  ${\it\Gamma}_{0}$ correspond with the energy of the system in the unstable state and its decay rate at the exponential  (or canonical) regime of the decay process. $E_{\text{min}}$ is the minimal (the lowest) energy of the system. Inserting $\omega_{BW}(E)$ into formula (\ref{a-spec}) for the amplitude $A(t)$ after some algebra one finds that
\begin{equation}
A(t) = \frac{N}{2\pi}\,
e^{\textstyle{ - \frac{i}{\hbar} E_{0}t }}\, I_{\beta}\left(\frac{{\it\Gamma}_{0} t}{\hbar}\right), \label{I(t)a}
\end{equation}
where
\begin{equation}
I_{\beta}(\tau) \stackrel{\rm def}{=}\int_{-\beta}^{\infty}
 \frac{1}{\eta^{2}
+ \frac{1}{4}}\, e^{\textstyle{ -i\eta\tau}}\,d\eta. \label{I(t)}
\end{equation}
Here $\tau = \frac{{\it\Gamma}_{0} t}{\hbar} \equiv \frac{t}{\tau_{0}}$, $\tau_{0}$ is the lifetime  and $\beta = \frac{E_{0} - E_{min}}{{\it\Gamma}_{0}}$. The integral $I_{\beta}(t)$ can be expressed in terms of special functions as follows
\begin{multline}
I_{\beta} (\tau) = 2\pi e^{\textstyle{-\frac{\tau}{2}}}\,+\,
\, i\Big\{e^{\textstyle{-\frac{\tau}{2}}}\,E_{1}\Big(-i(\beta-\frac{i}{2})\tau\Big) \\
- e^{\textstyle{+\frac{\tau}{2}}}\,E_{1}\Big(-i(\beta+ \frac{i}{2})\tau\Big)\Big\}, \label{I(t)-end1}
\end{multline}
where $E_{1}(z)$ denotes the integral--exponential function defined according to \cite{olver,abramowitz}, ($z$ is a complex number).

Next using this $A(t)$ given  by relations (\ref{I(t)a}), (\ref{I(t)}) and the relation (\ref{h(t)}) defining the  effective Hamiltonian $h_{\phi}(t)$ one finds that within the Breit-Wigner model considered
\begin{equation}
h(t) = i \hbar \frac{1}{A(t)}\,\frac{\partial A(t)}{\partial t} = E_{0} + {\it\Gamma}_{0}\,\frac{J_{\beta}(\frac{{\it\Gamma}_{0} t}{\hbar})}{I_{\beta}(\frac{{\it\Gamma}_{0} t}{\hbar})}, \label{h(t)-1}
\end{equation}
where
\begin{equation}
J_{\beta}(\tau) = \int_{- \beta}^{\infty}\,\frac{x}{x^{2} + \frac{1}{4}}\,e^{\textstyle{-ix\tau}}\,dx. \label{J-R}
\end{equation}
It is important to be aware of the following problem: Namely from the definition of $J_{\beta}(\tau)$  one can conclude that $J_{\beta}(0)$ is undefined ($\lim_{\tau \to 0} \,J_{\beta}(\tau) = \infty$). This is because within the model defined by the Breit-Wigner distribution of the energy density, $\omega_{BW}(E)$, the expectation value of $H$, that is $\langle \phi|H|\phi \rangle $ is not finite. So all the consideration based on the use of $J_{\beta}(\tau)$ are valid only for $\tau > 0$.

Note that simply
\begin{equation}
J_{\beta}(\tau) \equiv i\frac{\partial I_{\beta} (\tau)}{\partial \tau} , \label{J-R-eq}
\end{equation}
which allows one to find analytical form of  $J_{\beta}(\tau)$ having such a form for $I_{\beta}(\tau)$.

We need to know the energy of the system in the unstable state $|\phi\rangle$ considered. The instantaneous energy $E_{\phi}(t)$ of the system in the unstable state $|\phi\rangle$  is given by the relation (\ref{m(t)}). So within the Breit-Wigner model one finds that
\begin{equation}
E_{\phi}(t)  = E_{0} + {\it\Gamma}_{0}\,\Re\,\left[\frac{J_{\beta}(\frac{{\it\Gamma}_{0} t}{\hbar})}{I_{\beta}(\frac{{\it\Gamma}_{0} t}{\hbar})}\right], \label{h(t)-2}
\end{equation}
or, equivalently
\begin{equation}
\kappa(t) \stackrel{\text{def}}{=}
\frac{E_{\phi}(t)  - E_{\text{min}}}{E_{0}  - E_{\text{min}}} = 1\,+\,\frac{1}{\beta}\,\Re\,\left[\frac{J_{\beta}(\frac{{\it\Gamma}_{0} t}{\hbar})}{I_{\beta}(\frac{{\it\Gamma}_{0} t}{\hbar})}\right]. \label{h(t)-2a}
\end{equation}
(This relation, i.e. $\kappa (t)$, was studied, for example in \cite{Urbanowski:2016pks,Urbanowski:2013tfa}).

It is relatively simple to find asymptotic expressions $I_{\beta}{\tau}$ and $J_{\beta}(\tau)$ for $\tau \to \infty$ directly from (\ref{I(t)}) and (\ref{J-R}) using , e.g., method of the integration by parts. We have for $\tau \to \infty$:
\begin{multline}
I_{\beta}(\tau) \simeq \frac{i}{\tau}\,\frac{e^{\textstyle{i\beta \tau}}}{\beta^{2} + \frac{1}{4}}\,\left\{-1 + \frac{2 \beta}{\beta^{2} + \frac{1}{4}}\,\frac{i}{\tau} \right. \\ \left. + \left[\frac{2}{\beta^{2} + \frac{1}{4}} - \frac{8 \beta^{2}}{(\beta^{2} + \frac{1}{4})^{2}}\right]\,\left(\frac{i}{\tau}\right)^{2} + \ldots \right\} \label{I-as}
\end{multline}
and
\begin{multline}
J_{\beta}(\tau) \simeq \frac{i}{\tau}\,\frac{e^{\textstyle{i\beta \tau}}}{\beta^{2} + \frac{1}{4}}\,\left\{\beta  +
\left[1 - \frac{2 \beta^{2}}{\beta^{2} + \frac{1}{4}}\right]\,\frac{i}{\tau} \right. \\ \left. + \frac{\beta}{\beta^{2} + \frac{1}{4}}\left[\frac{8 \beta^{2}}{\beta^{2} + \frac{1}{4}} - 6\right]\,\left(\frac{i}{\tau}\right)^{2} + \ldots \right\}. \label{J-as}
\end{multline}
These two last asymptotic expressions alows one to find  for $\tau \to \infty$ the asymptotic form of the ratio $\frac{J_{\beta}(\tau)}{I_{\beta}(\tau)}$ used in relations
(\ref{h(t)-1}), (\ref{h(t)-2}) and (\ref{h(t)-2a}), which has much simpler form than asymptotic expansions for $I_{\beta}(\tau)$ and $J_{\beta}(\tau)$. One finds that for $\tau \to \infty$,
\begin{equation}
\frac{J_{\beta}(\tau)}{I_{\beta}(\tau)} \,\simeq \, - \,\beta \;-\;\frac{i}{\tau}\;-\; \frac{2 \beta}{\beta^{2} + \frac{1}{4}}\,\frac{1}{\tau^{2}} \,\,+ \ldots .\label{J-I-as}
\end{equation}
Starting from this asymptotic expression and formula (\ref{h(t)-2}) or
making use of the asymptotic expansion of $E_{1}(z)$ \cite{abramowitz} and (\ref{I(t)-end1}),
\begin{equation}
{E_{1}(z)\vline}_{\, |z| \rightarrow \infty} \sim
\frac{e^{\textstyle{ -z}}}{z} \left( 1 - \frac{1}{z} + \frac{2}{z^{2}} -
\ldots \right),  \label{E1-as}
\end{equation}
where $| \arg z  | < \frac{3}{2} \pi$,
one finds, eg. that for $t \to \infty$,
\begin{equation}
{E_{\phi}(t)\vline}_{\,t \rightarrow \infty} \simeq  { E}_{\text{min}}\, -\,2\,
\frac{ { E}_{0}\,-\,{E}_{min}}{ |\,h_{\phi}^{0}\,-\,{E}_{\text{min}} \,|^{\,2} }  \;
\left(\frac{\hbar}{t} \right)^{2} ,\label{Re-h-as}
\end{equation}
where $h_{\phi}^{0} = E_{0} - \frac{i}{2}{\it\Gamma}_{0}$. This last relation is valid for $t > T$, where $T$ denotes the cross--over time, ie. the time when
exponential and late time inverse power law contributions to the survival amplitude begin to be comparable.

\begin{figure}[h!]
\begin{center}
\includegraphics[width=0.48\textwidth]{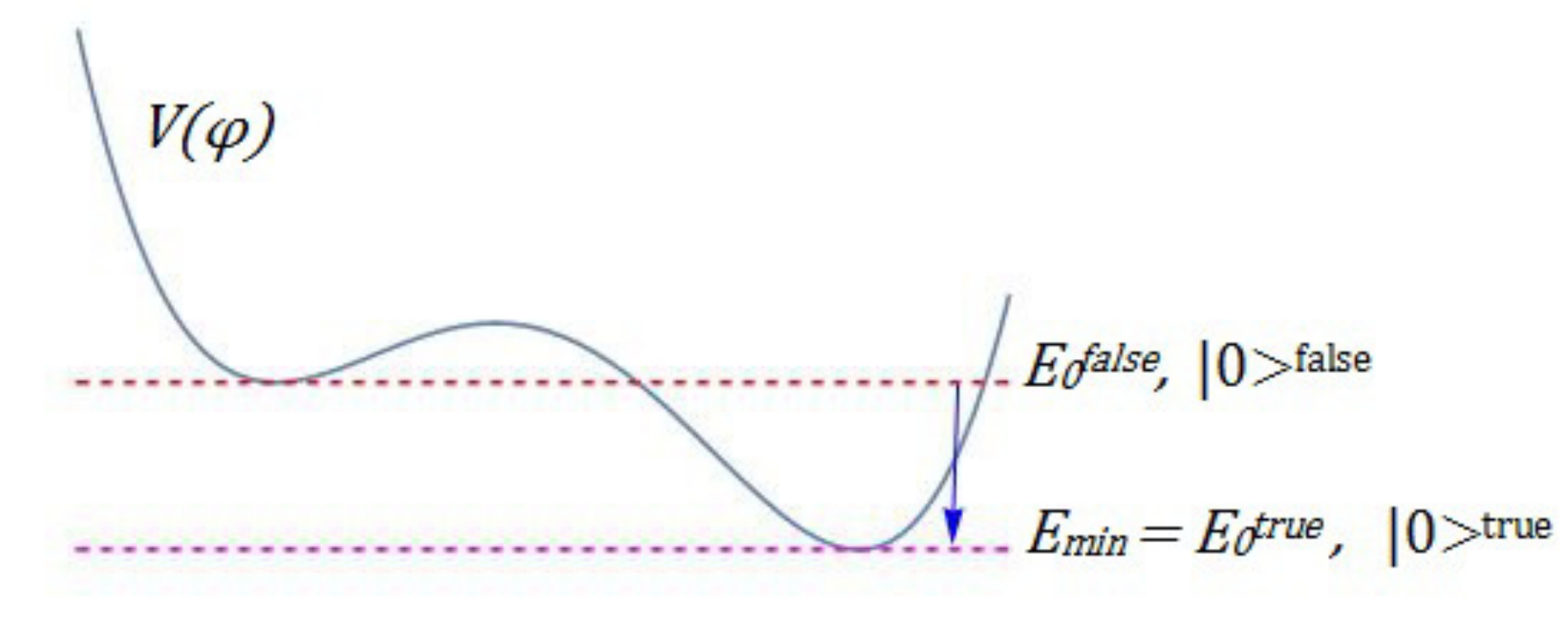}\\
\caption{Transition of the system from the false vacuum state $|0\rangle^{\text{false}}$ to the true ground state of the system, i.e. the true vacuum state $|0\rangle^{\text{true}}$.
States $|0\rangle^{\text{false}}$ and $|0\rangle^{\text{true}}$ correspond to the local minimum  and to the true lowest minimum of the potential $V(\varphi)$ of the scalar field $\varphi$ respectively.
}
  \label{fig:fig15}
\end{center}
\end{figure}

Some cosmological scenario predict the possibility of decay of the Standard Model vacuum at an inflationary stage of the evolution of the universe (see eg. \cite{Branchina:2013jra} and also \cite{Bezrukov:2015ytd} and reference therein) or earlier. Of course this decaying Standard Model vacuum is described by the quantum state corresponding to a local minimum of the energy density which is not the absolute minimum of the energy density of the system considered (see, eg. Fig.~\ref{fig:fig15}).
The scenario in which false vacuum may decay at the inflationary stage of the universe corresponds with the hypothesis analyzed by Krauss and Dent \cite{Krauss:2007rx,Krauss:2008pt}.
Namely in the mentioned papers the hypothesis
that some false vacuum regions do survive well
up to the cross--over time $T$ or  later was considered  where $T$ is the same cross-over time which is is considered within
the  theory of evolving in time quantum unstable systems.
The fact that the decay of the false vacuum  is the quantum decay process means that state vector corresponding  to the false vacuum is a quantum unstable (or metastable) state. Therefore all the general properties of quantum unstable systems  must also occur
 in the case of such a quantum unstable state as the false vacuum.
This applies in particular to such properties as late time deviations from the exponential decay law and properties of the energy $E^{\text{false}}_{0}(t)$  of the
system in the quantum false vacuum  state at late times $t > T$. In \cite{Urbanowski:2011zz} it was pointed out the energy of those false vacuum regions which survived up to $T$ and much later differs from $E^{\text{false}}_{0}$ \cite{Urbanowski:2011zz}.

So within the  cosmological scenario in which the decay of false vacuum is assumed the unstable state $|\phi\rangle$ corresponds with the false vacuum state: $|\phi\rangle = |0\rangle^{\text{false}}$. Then $|0\rangle^{\text{true}}$ is the true vacuum state, that is the state corresponding to the true minimal energy. In such a case  $E_{0} \rightarrow E_{0}^{\text{false}}$ is the energy of a state corresponding to the false vacuum measured at the canonical decay time (the exponential decay regime) and $E^{\text{true}}_{0}$ is the energy of true vacuum (i.e., the true ground state of the system), so $E^{\text{true}}_{0} \equiv E_{\text{min}}$.
The corresponding quantum mechanical process looks as it is shown in Fig.~\ref{fig:fig15}.
If one wants to generalize the above results obtained on the basis of quantum mechanics to quantum field theory one should take into account among others a volume factors so that survival probabilities per unit volume  should be considered and similarly the energies and the decay rate: $E \mapsto  \rho (E) = \frac{E}{V_{0}}$, ${\it\Gamma}_{0} \mapsto \gamma = \frac{{\it\Gamma}_{0}}{V_{0}}$, where $V_{0} = V(t_{0})$ is the volume of the considered system at the initial instant $t_{0}$, when the time evolution starts.
The volume $V_{0}$ is used in these considerations because
the initial unstable state $|\phi\rangle \equiv |0\rangle^{\text{false}}$ at $t=t_{0}=0$ is expanded into eigenvectors $|E\rangle$ of $H$ at this initial instant $t_{0}$, (where $E \in \sigma_{c}(H)$)  and then this expansion is used to find the density of the energy distribution $\omega (E)$. It is easy to see that the mentioned changes
$E \mapsto  \frac{E}{V_{0}}$ and  ${\it\Gamma}_{0} \mapsto \frac{{\it\Gamma}_{0}}{V_{0}}$ do not changes integrals $I_{\beta}(t)$ and $J_{\beta}(t)$ and the relation (\ref{h(t)-2a}).
Similarly in such a situation the parameter $\beta = \frac{E_{0} - E_{min}}{{\it\Gamma_{0}}}$ does not changes.
This means that the relations (\ref{h(t)-2}), (\ref{h(t)-2a}), (\ref{Re-h-as}) can be replaced by corresponding relations for the densities $\rho_{\text{de}}$ or $\Lambda$ (see, eg., \cite{Urbanowski:2016pks,Urbanowski:2012pka,ms-ku2}). Within such an approach $E(t)$ corresponds to the running cosmological constant $\Lambda(t)$ and $E_{\text{min}}$ to the $\Lambda_{bare}$. The parametrization used in next Sections is based on relations (\ref{h(t)-2}), (\ref{h(t)-2a}). Integrals (\ref{J1}), (\ref{I1}) introduced in Sec. 1 are obtained from (\ref{J-R}) and (\ref{I(t)}) replacing $\beta$ by $\frac{1 - \alpha}{\alpha}$. Similarly solutions (\ref{J11}) and (\ref{I11}) correspond to (\ref{I(t)-end1}) and to the function $J_{\beta}(\tau)$ obtained from (\ref{I(t)-end1}) using (\ref{J-R-eq}).

\section{Cosmological equations with $\rho_\text{de}=\Lambda_\text{bare}+E_\text{R}\left[1+\frac{\alpha}{1-\alpha} \Re\left(\frac{J(t)}{I(t)}\right)\right]$}

The cosmological model with the parametrization of the dark energy (\ref{param}) belonging to the class of parametrizations proposed in \cite{Urbanowski:2016pks} after putting $E_R = E_0 - \Lambda_{\text{bare}}$ assumes the following form of $\rho_\text{de}$ (we use units $8\pi G = c = 1$)
\begin{equation}
\rho_\text{de}=\Lambda_\text{bare}+E_\text{R}\left[1+\frac{\alpha}{1-\alpha} \Re\left(\frac{J(t)}{I(t)}\right)\right]. \label{parametrization}
\end{equation}
It can be introduced as the covariant theory from the following action
\begin{equation}
S=\int \sqrt{-g}(R+\mathcal{L}_\text{m}) \, d^4 x,\label{action}
\end{equation}
where $R$ is the Ricci scalar, $\mathcal{L}_\text{m}$ is the Lagrangian for the barotropic fluid and $g_{\mu \nu}$ is the metric tensor. We assume the signature of the metric tensor as $(+,-,-,-)$ and, for the simplicity, that the constant curvature is zero (the flat model). The Ricci scalar for the Friedmann-Lemaitre-Robertson-Walker (FLRW) metric is presented by the following formula
\begin{equation}
R=-6\left[\frac{\ddot a}{a}+\left(\frac{\dot a}{a}\right)^2\right]
\end{equation}
where a dot means the differentiation with respect to the cosmological time $t$.

The Lagrangian for the barotropic fluid is expressed by the formula
\begin{equation}
\mathcal{L}_\text{m}=-\rho_\text{tot}\left(1+\int\frac{p_\text{tot}(\rho_\text{tot})}{\rho_\text{tot}^2} \, d\rho_\text{tot}\right),
\end{equation}
where $\rho_\text{tot}$ is the total density of fluid and $p_\text{tot}(\rho_\text{tot})$ is the total pressure of fluid \cite{Minazzoli:2012md}. We assume that this fluid consists of three components: the baryonic matter $\rho_\text{b}$, the dark matter $\rho_\text{dm}$ and the dark energy $\rho_\text{de}$. We treat the baryonic matter and the dark matter like dust. In consequence the equations of state for them are following: $p_\text{b}(\rho_\text{b})=0$ and $p_\text{dm}(\rho_\text{dm})=0$. The equation of state for the dark energy is assumed in the form $p_\text{de}(\rho_\text{de})=-\rho_\text{de}$.

Of course, the total density is expressed by $\rho_\text{tot}=\rho_\text{b}+\rho_\text{dm}+\rho_\text{de}$ and the total pressure is expressed by $p_\text{tot}(\rho_\text{tot})=p_\text{de}(\rho_\text{de})=-\rho_\text{de}$.

We can find the Einstein equations using calculus of variations method by variation action (\ref{action}) by the metric $g_{\mu \nu}$. Then we get two equations: the Friedmann equation
\begin{equation}
3 H^2=3 \frac{\dot a}{a}^2=\rho_\text{tot}=\rho_\text{b}+\rho_\text{dm}+\rho_\text{de},\label{friedmann}
\end{equation}
where $H=\frac{\dot a}{a}$ is the Hubble function,
and the acceleration equation
\begin{equation}
\frac{\ddot a}{a}=-\frac{1}{6}(\rho_\text{tot}+3 p_\text{tot}(\rho_\text{tot}))=\rho_\text{b}+\rho_\text{dm}-2\rho_\text{de}.\label{acceleration}
\end{equation}

From Eqs.~(\ref{friedmann}) and (\ref{acceleration}) we can get the conservation equation
\begin{equation}
\dot\rho_\text{tot}=-3H (\rho_\text{tot}+p_\text{tot}(\rho_\text{tot})).
\end{equation}
The above equation can be rewritten as
\begin{equation}
\dot\rho_\text{m}=-3H \rho_\text{m}-\dot\rho_\text{de},\label{conservation}
\end{equation}
where $\rho_\text{m}=\rho_\text{b}+\rho_\text{dm}$.

Let $Q$ is the interaction between the dark matter and the dark energy. Then Eq.~(\ref{conservation}) is equivalent the following equations
\begin{align}
\dot\rho_\text{b} &= -3H \rho_\text{b}, \\
\dot\rho_\text{dm} &= -3H \rho_\text{dm}+Q
\end{align}
and
\begin{equation}
\dot\rho_\text{de}=-Q,\label{darkenergy}
\end{equation}
where the interaction $Q$ is defined by Eq.~(\ref{darkenergy}). The interaction between the dark matter and the dark energy can be interpreted as the energy transfer in the dark sector. If $Q>0$ then the energy flow is from the dark energy to the dark matter. If $Q<0$ then the energy flow is from the dark matter to the dark energy.

\begin{figure}
	\centering
	\includegraphics[width=0.45\textwidth]{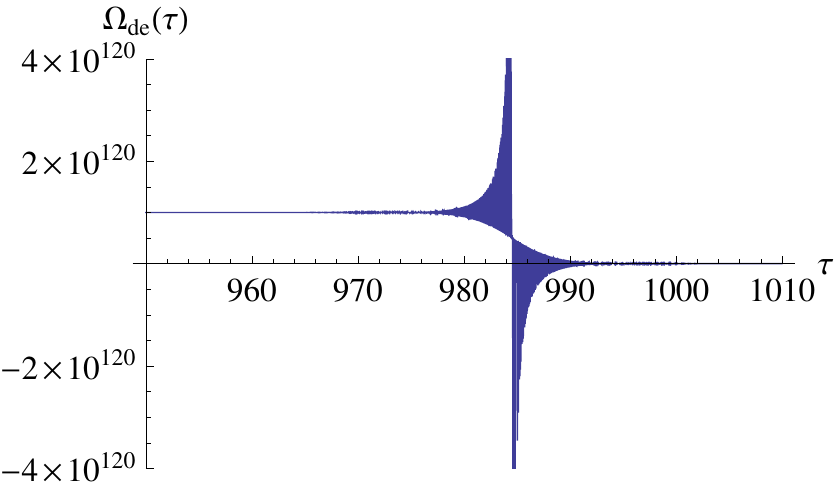}
	\caption{The dependence $\Omega_\text{de}(\tau)$ for $\alpha=10^{-105}$ and $\frac{E_0}{3H_0^2}=10^{120}$. The rescaled time $\tau$ is given in unit $[1.3 \times 10^{-40} s]$.}
	\label{fig:fig1}
\end{figure}

\begin{figure}
	\centering
	\includegraphics[width=0.45\textwidth]{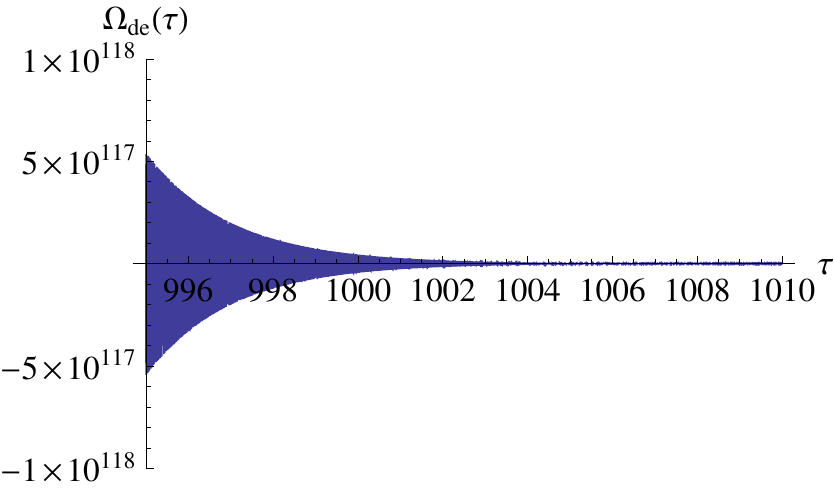}
	\caption{The dependence $\Omega_\text{de}(\tau)$ during the intermediate phase of damped oscillations for $\alpha=10^{-105}$ and $\frac{E_0}{3H_0^2}=10^{120}$. The rescaled time $\tau$ is given in unit $[1.3 \times 10^{-40} s]$.}
	\label{fig:fig2}
\end{figure}

For the description of the evolution of the universe is necessary to use the Friedmann equation (\ref{friedmann}) and the conservation equation (\ref{conservation}). These formulas can be rewritten in dimensionless parameters. Let $\Omega_\text{m}=\frac{\rho_\text{m}}{3H_0^2}$ and $\Omega_\text{de}=\frac{\rho_\text{de}}{3H_0^2}$, where $H_0$ is the present value of the Hubble function. Then from Eqs.~(\ref{friedmann}) and (\ref{conservation}), we get
\begin{equation}
\frac{H^2}{H_0^2}=\Omega_\text{m}+\Omega_\text{de}
\end{equation}
and
\begin{equation}
\dot\Omega_\text{m}=-3H \Omega_\text{m}-\dot\Omega_\text{de}.
\end{equation}
The above equations are sufficient to find the behavior of the matter, the dark energy, the Hubble function and the scale factor as a function of the cosmological time. We cannot find the exact solutions because these equations are too complicated. In this case we should search  for numerical solutions. The behavior of the dark energy, is presented in Figs.~\ref{fig:fig1} and \ref{fig:fig2}. Fig.~\ref{fig:fig1} shows the diagram of the dependence $\Omega_\text{de}(\tau)$ with respect of the rescaled time $\tau$ for $\alpha = 10^{-105}$ and $\frac{E_0}{3H_0^2} = 10^{120}$. On the diagram we can see the beginning value of the dark energy density, which is equal $\Omega_\text{de}\approx10^{120}$, is reduced to the present value of the dark energy density, which is $\Omega_\text{de}\approx 0.7$. This final value of $\Omega_{\text{de}}$ does not depend on the values of parameters $\alpha$ and $\frac{E_0}{3H_0^2}$. Therefore, this mechanism makes an attempt the cosmological constant problem. For the late time dark energy can be treated as the cosmological constant. The characteristic of the intermidiate oscillatory regime is depending on the parameter $\alpha$. With the increasing value of $\alpha$ the number of oscillations, their amplitude, their period as well as the length of this regime decreases. If $\alpha>0.4$ then oscillations begin to disappear and the value of $\Omega_{\text{de}}$ jumps to the constant value of $0.7$.

\begin{figure}
	\centering
	\includegraphics[width=0.45\textwidth]{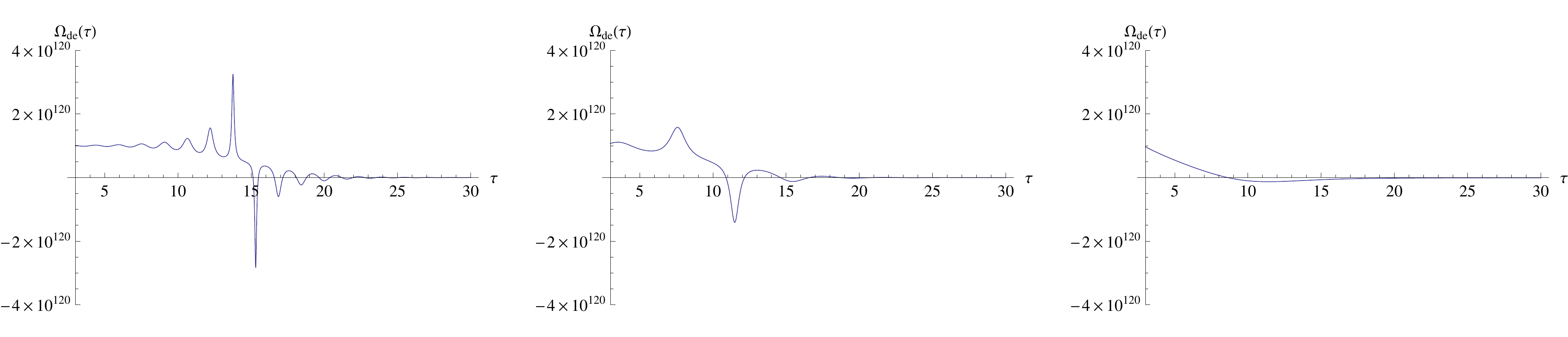}
	\caption{The dependence $\Omega_\text{de}(\tau)$ for $\alpha=0.2$ (left figure) and $\alpha=0.4$ (medium figure) and $\alpha=0.8$ (right figure) and $\frac{E_0}{3H_0^2}=10^{120}$. The rescaled time $\tau$ for the left figure is given in unit $[5.3 \times 10^{-145} s]$, for the center figure is given in unit $[2.0 \times 10^{-145} s]$ and for the right figure is given in unit $[3.3 \times 10^{-146} s]$}
	\label{fig:fig11}
\end{figure}

Fig.~\ref{fig:fig2} shows the diagram of the dependence $\Omega_\text{de}(\tau)$ during the intermediate phase of damped oscillations with respect of the time $\tau$ for $\alpha=10^{-105}$ and $\frac{E_0}{3H_0^2}=10^{120}$. Note that the dark energy oscillates and the amplitude of oscillations decreases with the time. In consequence the dark energy can be treated as the cosmological constant after the intermediate phase of oscillations. Fig.~\ref{fig:fig11} shows the diagrams of the dependence $\Omega_\text{de}(\tau)$ with respect of the time $\tau$ for different values of $\alpha$ ($\alpha=0.2,\text{ }0.4,\text{ }0.8$) and $\frac{E_0}{3H_0^2}=10^{120}$. This figure presents how the evolution of $\Omega_\text{de}(\tau)$ is depended on $\alpha$ parameter. Note that the oscillations disappear for $\alpha>0.4$.

In general, if $\alpha$ decreases then  times when oscillatory regime takes place increase. This means that passage from the very high energies to the extremely small energies, which takes place at the oscillatory regime, moves in the direction of increasing time with decreasing $\alpha$  and for suitable small value of $\alpha$ this oscillatory regime can occur at relatively late times.

\begin{figure}
	\centering
	\includegraphics[width=0.45\textwidth]{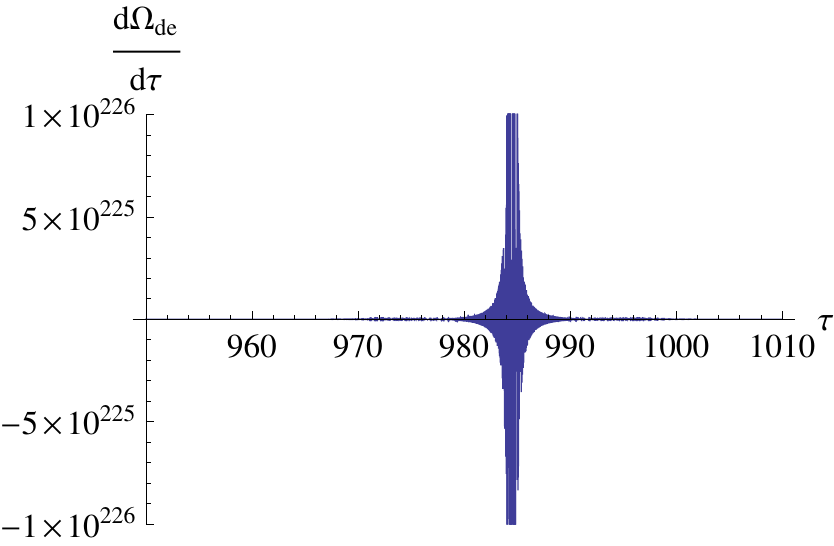}
	\caption{The dependence $\frac{d\Omega_\text{de}}{d\tau} (\tau)$ for $\alpha=10^{-105}$ and $\frac{E_0}{3H_0^2}=10^{120}$. Note that for the negative value of $\frac{d\Omega_\text{de}}{d\tau}$, the energy is transfered from the dark energy to the dark matter and for the positive value of $\frac{d\Omega_\text{de}}{d\tau}$, the energy is transfered from the dark matter to the dark energy. The rescaled time $\tau$ is given in unit $[1.3 \times 10^{-40} s]$.}
	\label{fig:fig3}
\end{figure}

\begin{figure}
	\centering
	\includegraphics[width=0.45\textwidth]{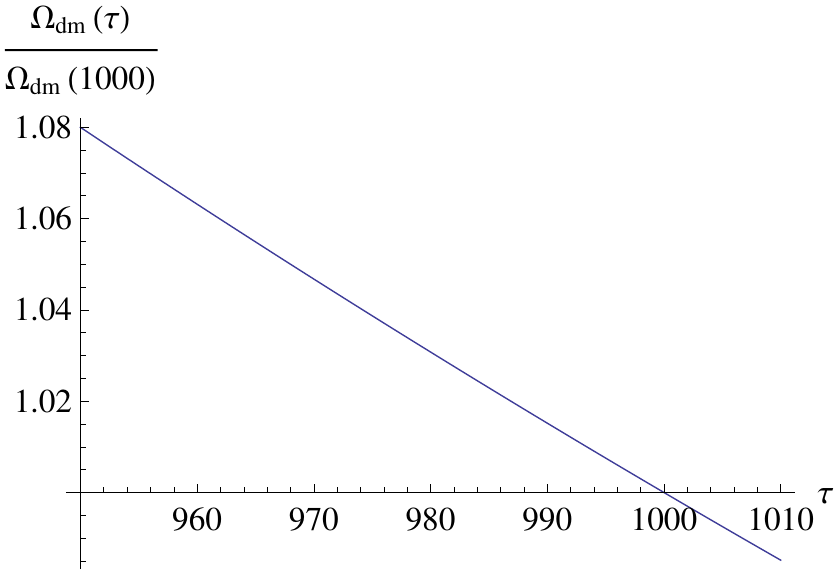}
	\caption{The dependence $\Omega_\text{dm}$ for $\alpha=10^{-105}$ and $\frac{E_0}{3H_0^2}=10^{120}$. We include influence of the radiation for the evolution of the matter. Note that the dark energy has the negligible influence for the evolution of the matter. The rescaled time $\tau$ is given in unit $[1.3 \times 10^{-40} s]$.}
	\label{fig:fig4}
\end{figure}

\begin{figure}
	\centering
	\includegraphics[width=0.45\textwidth]{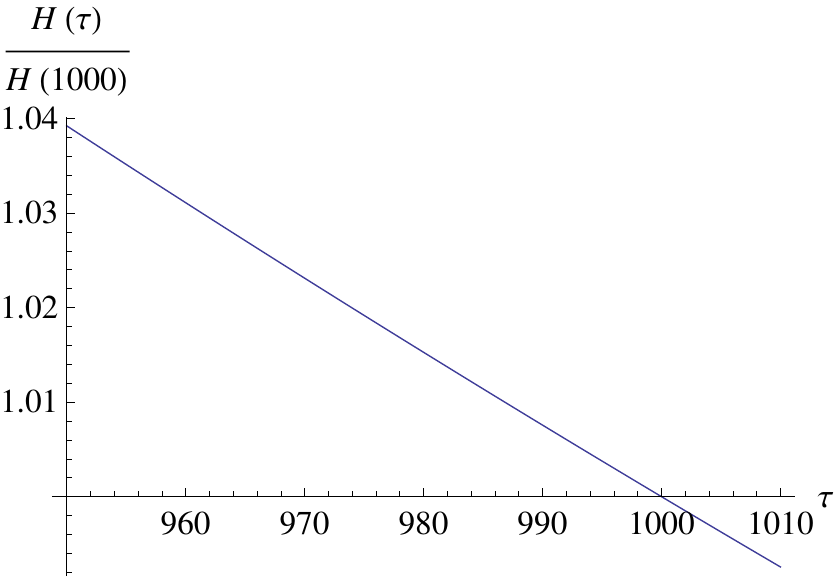}
	\caption{The dependence $H(\tau)$ for $\alpha=10^{-105}$ and $\frac{E_0}{3H_0^2}=10^{120}$. We include influence of the radiation for the evolution of the Hubble function. Note that dark energy has the negligible influence for the evolution of the Hubble function. The rescaled time $\tau$ is given in unit $[1.3 \times 10^{-40} s]$.}
	\label{fig:fig5}
\end{figure}

\begin{figure}
	\centering
	\includegraphics[width=0.45\textwidth]{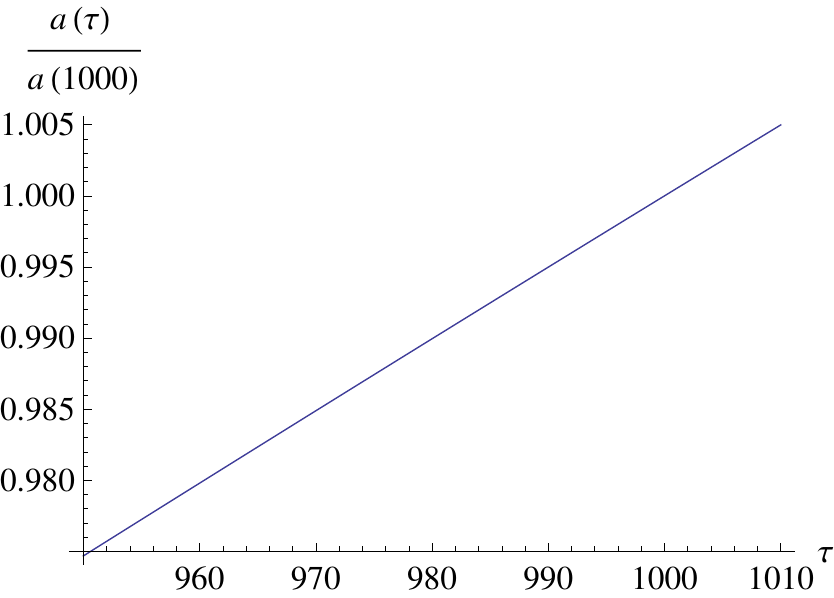}
	\caption{The dependence $a(\tau)$ for $\alpha=10^{-105}$ and $\frac{E_0}{3H_0^2}=10^{120}$. We include the influence of the radiation for the evolution of the scale factor. Note that dark energy has the negligible influence for the evolution of the scale factor. The rescaled time $\tau$ is given in unit $[1.3 \times 10^{-40} s]$.}
	\label{fig:fig10}
\end{figure}

Figure \ref{fig:fig3} presents the evolution of $\frac{d\Omega_\text{de}}{d\tau}$. The evolution of the matter is demonstrated in Fig.~\ref{fig:fig4} and the Hubble function, is presented in Fig.~\ref{fig:fig5}. The diagram of the scale factor with respect to the cosmological time is presented in Fig.~\ref{fig:fig10}. 


We have
$\tau = \frac{\alpha(E_0-\Lambda_\text{bare})}{\hbar (1-\alpha)}V_0 t$
therefore if the value of the parameter $\alpha$ increases than the damping of oscillations should be also increased. In the in the limiting case, if $\alpha$ is equal zero then we get the $\Lambda$CDM model. This last conclusion can be easily drawn analyzing the late time properties of $\rho_\text{de}$.

For the late time, $\tau \to \infty$,  according to  the relation (\ref{J-I-as}), the parametrization of dark energy (\ref{parametrization}) can approximated by the following expression

\begin{equation}
\rho_\text{de}=\Lambda_\text{bare}-2 E_\text{R}\, \frac{\alpha^{2}}{(1 - \alpha)^{2} + \frac{\alpha^{2}}{4}}\,\frac{1}{\tau^{2}} + \ldots\;.
\label{approximation}
\end{equation}
From this relation the important observation follows: For any $\alpha >0$ the $\Lambda$CDM model is the limiting case, when $\tau \to \infty$, of the our model. So for very, very late times results obtained within our model and within $\Lambda$CDM model have to coincide. This parametrization of the dark energy was considered in \cite{ms-ku2, Szydlowski:2015rga, Szydlowski:2015fya}.

The dark energy is significantly lower than the energy density of matter in the early universe, which has a consequence that the transfer to the dark sector is negligible (see Fig.~\ref{fig:fig3}). Our model makes an
attempt the cosmological constant problem. In general, the amplitude of oscillations
of the dark energy decreases with time.

Thus for the late time universe, oscillations are negligible and the dark energy has the form of the cosmological constant.

\begin{figure}
	\centering
	\includegraphics[width=0.45\textwidth]{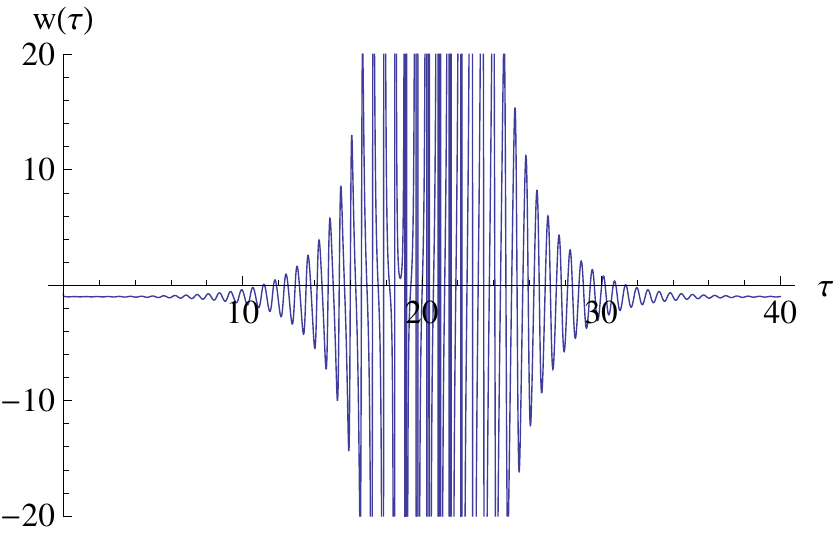}
	\caption{The typical dependence $w(\tau)$. This example is for $\alpha=0.09$ and $\frac{E_0}{3H_0^2}=10^{120}$. Note that after the intermediate phase of oscillations, the function $w(\tau)$ can be treated as a constant, which is equal $-1$. The rescaled time $\tau$ is given in unit $[1.3 \times 10^{-144} s]$.}
	\label{fig:fig9}
\end{figure}

The conservation equation for the dark energy (\ref{darkenergy}) can be rewritten as
\begin{equation}
\dot\rho_\text{de}=-3H (\rho_\text{de}+p_\text{de}),
\end{equation}
where $p_\text{de}$ is an effective pressure of the dark energy. In this case the equation of state for the dark energy is expressed by the following formula
\begin{equation}
p_\text{de}=w(t)\rho_\text{de},
\end{equation}
where the function $w(t)$ is given by the expression
\begin{equation}
w(t)=-1-\frac{\dot\rho_\text{de}}{\sqrt{3}\sqrt{\rho_\text{m}+\rho_\text{de}}\rho_\text{de}}=-1-\frac{1}{3H}\frac{d\ln\rho_\text{de}}{dt}.
\end{equation}
The diagram of coefficient equation of state $w(t)$ is presented in Fig.~\ref{fig:fig9}. The function $w(t)$, for the late time, is a constant and equal $-1$ which means that it describes the cosmlogical constant parameter. Note that the function $w(t)$ is also equal $-1$ which means that $\rho_{de}$ is constant as a consequance of the conservation condition (transfer between sectors is negligible). Therefore, the energy transfer is an effective process only during intermediate oscillation period (quantum regime).

\begin{figure}
	\centering
	\includegraphics[width=0.45\textwidth]{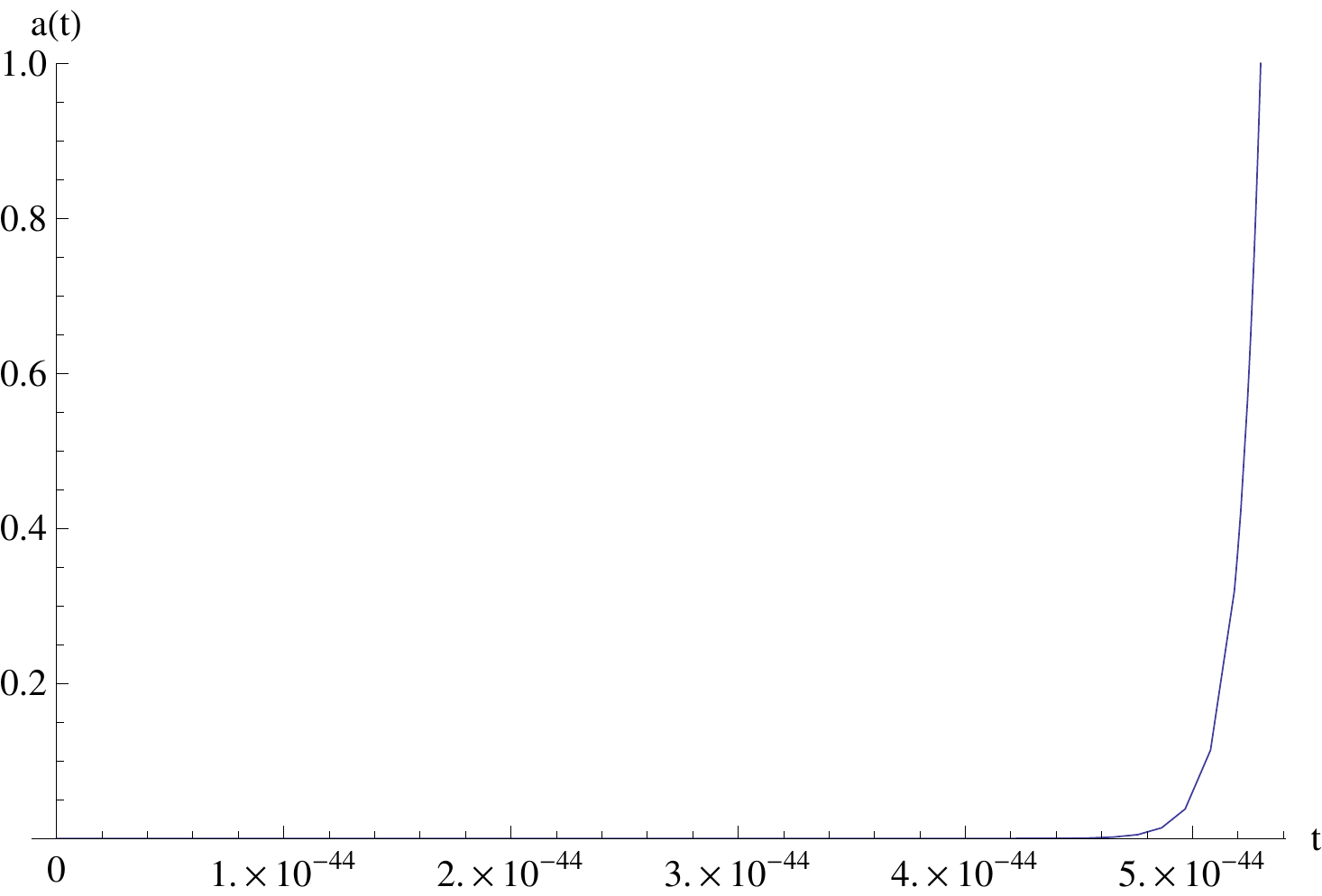}
	\caption{The dependence $a(t)$ for $\frac{E_0}{3H_0^2}=2*10^{125}$. We assume that $\rho_\text{de}\gg\rho_\text{m}$ and the intermediate phase of oscillations is after the Planck epoch. Note that, for the above assumptions, an inflation appears after the Planck epoch. The characteristic number of e-foldings of this inflation is equal 53 here. The cosmological time $t$ is given in seconds.}
	\label{fig:fig16}
\end{figure}

Let $\rho_\text{de} \gg \rho_\text{m}$. Then our model predicts an inflation. The formula for e-foldings $N=H_\text{init}(t_\text{fin}-t_\text{init})$ (see \cite{DeFelice:2010aj}) gets the following expression for our model
\begin{equation}
N=\sqrt{\frac{E_0}{3}}(t_\text{fin}-t_\text{init}),
\end{equation}
where $t_\text{init}\approx 0$ and $t_\text{fin}$ is the time of appearing of the intermediate phase of oscillations. Figure \ref{fig:fig16} presents the evolution of the scale factor $a$ with respect to the cosmological time during the inflation.

\section{Statistical analysis}

To estimate the model parameters we use the astronomical observations such as the supernovae of type Ia (SNIa), BAO, measurements of $H(z)$ for galaxies, the Alcock-Paczy{\'n}ski test and the measurements CMB.

The data of supernovae of type Ia, which were used in this paper, are taken from the Union 2.1 dataset \cite{Suzuki:2011hu}. In this context we use the following likelihood function
\begin{equation}
\ln L_{\text{SNIa}} = -\frac{1}{2} [A - B^2/C + \log(C/(2 \pi))],
\end{equation}
where $A=
(\mathbf{\mu}^{\text{obs}}-\mathbf{\mu}^{\text{th}})\mathbb{C}^{-1}(\mathbf{\mu}^{\text{obs}}-\mathbf{\mu}^{\text{th}})$,
$B=
\mathbb{C}^{-1}(\mathbf{\mu}^{\text{obs}}-\mathbf{\mu}^{\text{th}})$,
$C=\text{Tr} \mathbb{C}^{-1}$ and $\mathbb{C}$ is a covariance
matrix for SNIa. The observer distance modulus $\mu^{\text{obs}}$ is defined by the formula $\mu^{\text{obs}}=m-M$ (where $m$ is the apparent magnitude and
$M$ is the absolute magnitude of SNIa). The theoretical distance modulus is given by $\mu^{\text{th}} = 5\log_{10} D_L +25$ (where the luminosity distance is $D_L= c(1+z)\int_{0}^{z} \frac{d z'}{H(z)}$).

We use the following BAO data: Sloan Digital Sky Survey
Release 7 (SDSS DR7) dataset at $z = 0.275$
\cite{Percival:2009xn}, 6dF Galaxy Redshift Survey measurements at
redshift $z = 0.1$ \cite{Beutler:2011hx}, and WiggleZ
measurements at redshift $z = 0.44, 0.60, 0.73$
\cite{Blake:2012pj}. The likelihood function is defined by the
expression
\begin{equation}
\ln L_{\text{BAO}} = -
\frac{1}{2}\left(\mathbf{d}^{\text{obs}}-\frac{r_s(z_d)}{D_V(\mathbf{z})}\right)\mathbb{C}^{-1}\left(\mathbf{d}^{\text{obs}}-\frac{r_s(z_d)}{D_V(\mathbf{z})}\right),
\end{equation}
where $r_s(z_d)$ is the sound horizon at the drag epoch
\cite{Hu:1995en,Eisenstein:1997ik}.

Measurements of the Hubble parameter $H(z)$ of galaxies were taken from
\cite{Simon:2004tf,Stern:2009ep,Moresco:2012jh}. The likelihood function is given by the following formula
\begin{equation}\label{hz}
\ln L_{H(z)} = -\frac{1}{2} \sum_{i=1}^{N}  \left
(\frac{H(z_i)^{\text{obs}}-H(z_i)^{\text{th}}}{\sigma_i
}\right)^2.
\end{equation}

The likelihood function for the Alcock-Paczynski test
\cite{Alcock:1979mp,Lopez-Corredoira:2013lca} has the following form
\begin{equation}
\ln L_{AP} =  - \frac{1}{2} \sum_i \frac{\left(
AP^{th}(z_i)-AP^{obs}(z_i) \right)^2}{\sigma^2}.
\end{equation}
where $AP(z)^{\text{th}} \equiv \frac{H(z)}{z} \int_{0}^{z}
\frac{dz'}{H(z')}$ and $AP(z_i)^{\text{obs}}$ are observational
data
\cite{Sutter:2012tf,Blake:2011ep,Ross:2006me,Marinoni:2010yoa,daAngela:2005gk,Outram:2003ew,Anderson:2012sa,Paris:2012iw,Schneider:2010hm}.

\begin{figure}[ht]
    \centering
    \includegraphics[width=0.45\textwidth]{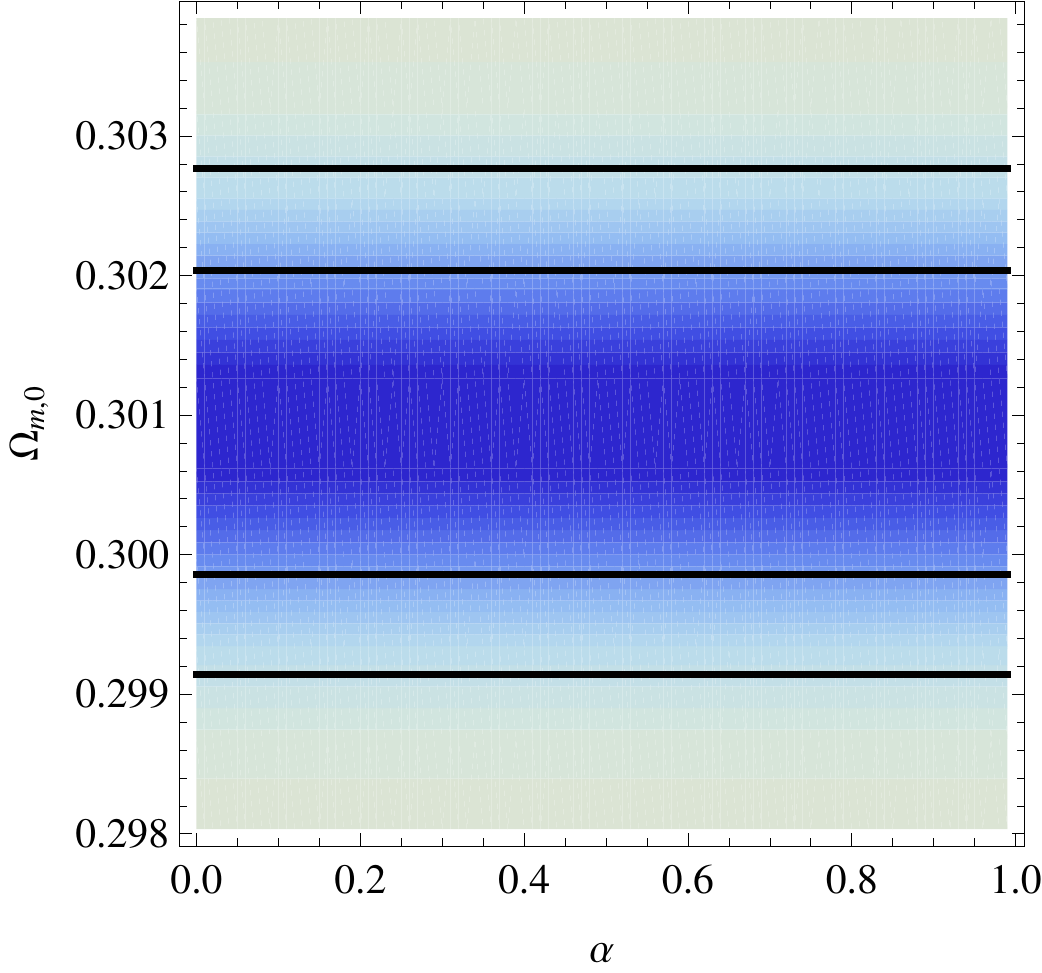}
    \caption{The intersection of the likelihood function of two model parameters ($\Omega_{\text{m},0}$, $\alpha$), with the marked $68\%$ and $95\%$ confidence levels. The plane of the intersection is the best fit of $H_0$ ($H_0=68.82 \left[\frac{\text{km}}{\text{s}\times\text{Mpc}}\right]$). We assumed that $E_0/(3H_0^2)$ is equal $10^{120}$, but changing of the value of $E_0/(3H_0^2)$ does not influence for results. Note that the values of the likelihood function are not sensitive to changing of $\alpha$ parameter.}
    \label{fig:fig6}
\end{figure}

\begin{figure}[ht]
    \centering
    \includegraphics[width=0.45\textwidth]{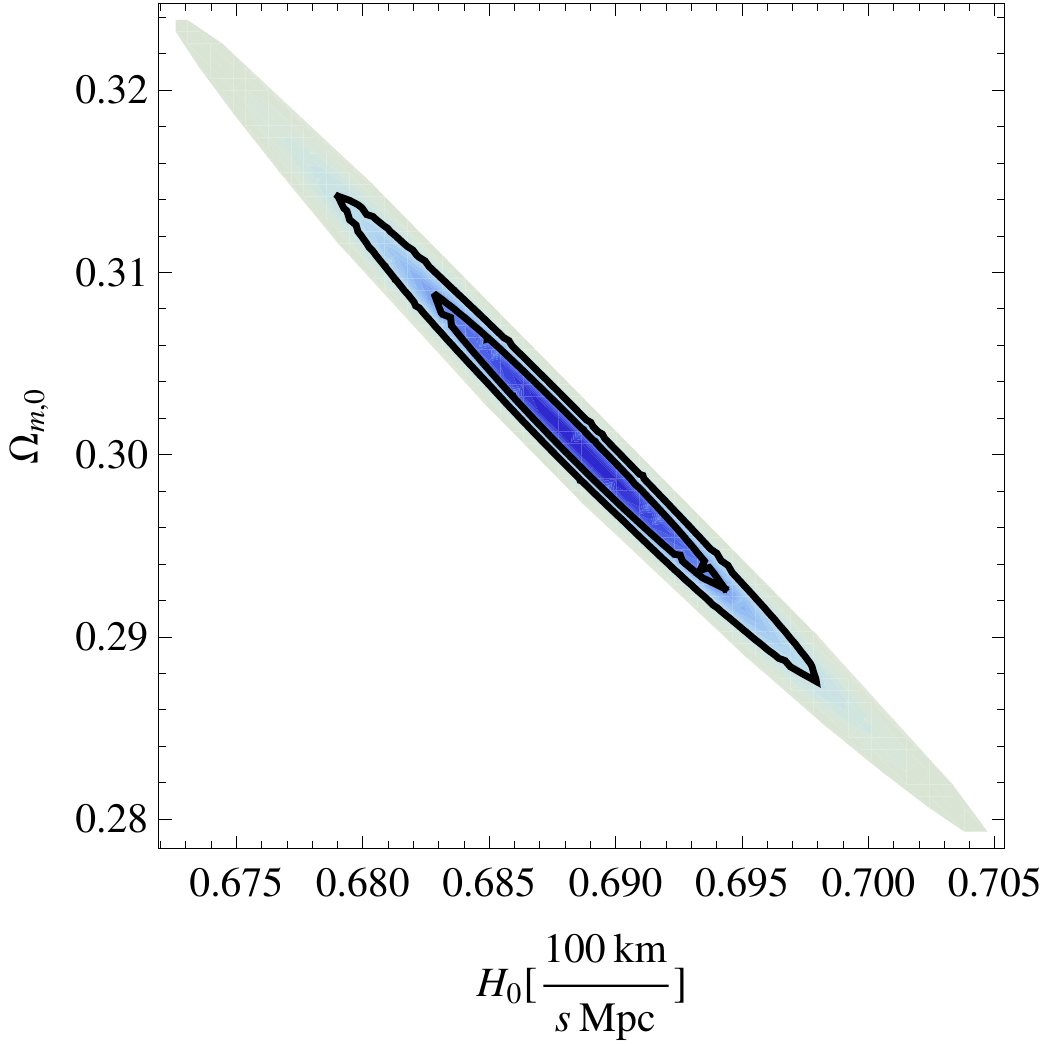}
    \caption{The intersection of the likelihood function of two model parameters ($\Omega_{\text{m},0}$, $H_0$), with the marked $68\%$ and $95\%$ confidence levels. The plane of the intersection is $\alpha=0.5$ and $E_0=10^{120}$.}
    \label{fig:fig7}
\end{figure}

\begin{figure}[ht]
    \centering
    \includegraphics[width=0.45\textwidth]{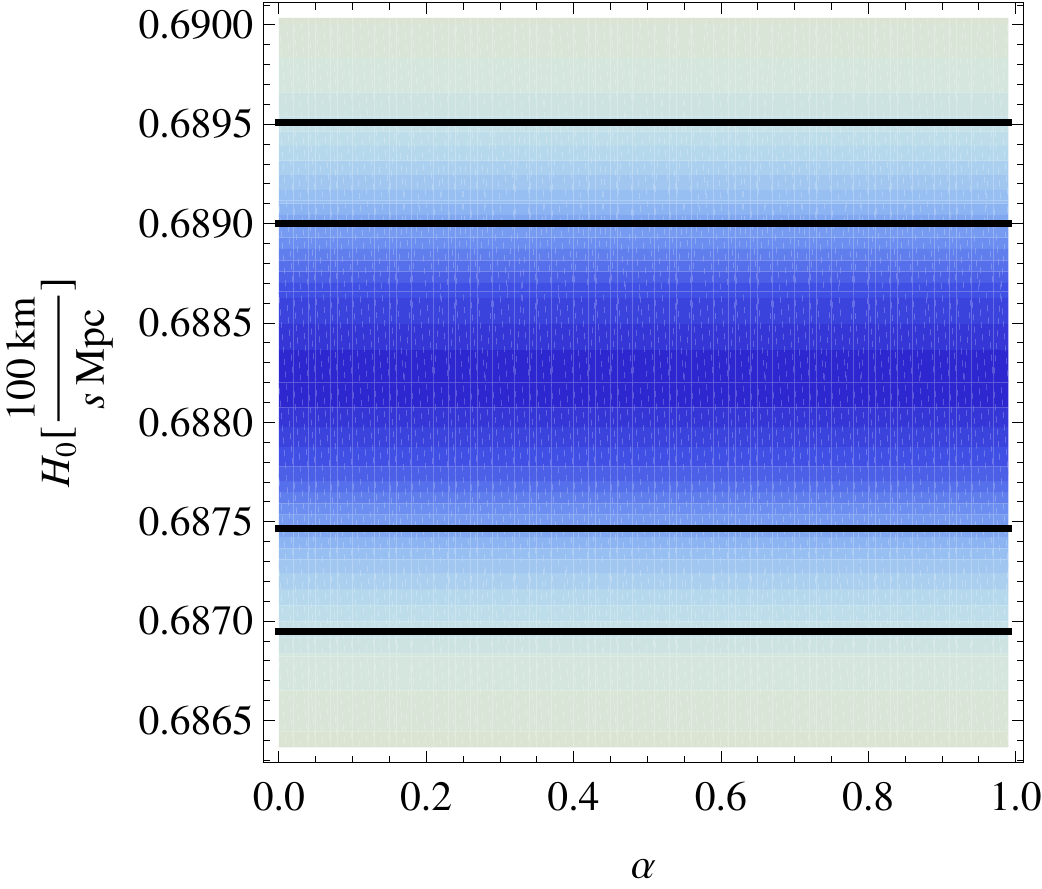}
    \caption{The intersection of the likelihood function of two model parameters ($H_0$, $\alpha$), with the marked $68\%$ and $95\%$ confidence levels. The plane of the intersection is the best fit of $\Omega_{\text{m},0}$ ($\Omega_{\text{m},0}=0.3009$). We assumed that $E_0/(3H_0^2)$ is equal $10^{120}$, but changing of the value of $E_0/(3H_0^2)$ does not influence for results. Note that the values of the likelihood function are not sensitive to changing of $\alpha$ parameter.}
    \label{fig:fig8}
\end{figure}

\begin{figure}[ht]
    \centering
    \includegraphics[width=0.45\textwidth]{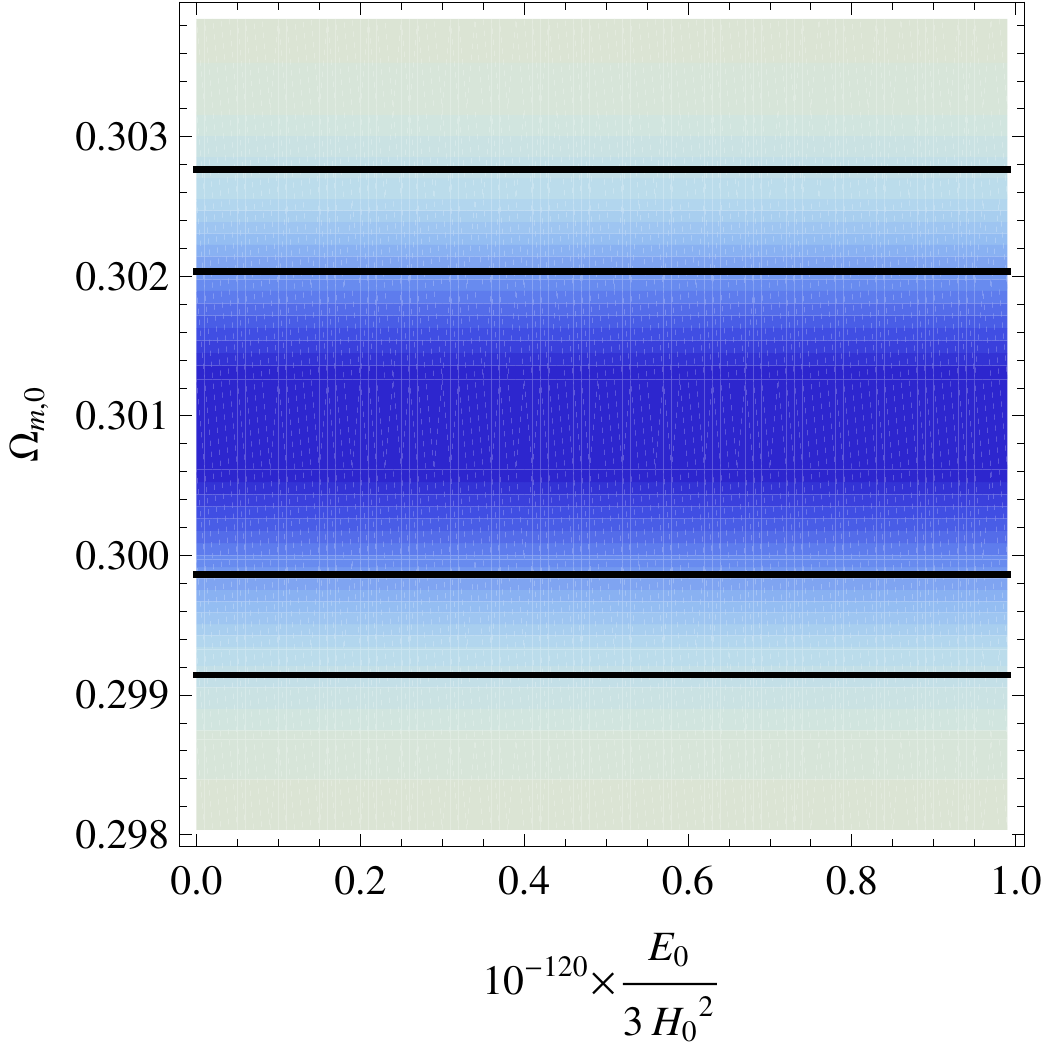}
    \caption{The intersection of the likelihood function of two model parameters ($\Omega_{\text{m},0}$, $\frac{E_0}{3H_0^2}$), with the marked $68\%$ and $95\%$ confidence levels. The plane of the intersection is the best fit of $H_0$ ($H_0=68.82 \left[\frac{\text{km}}{\text{s}\times\text{Mpc}}\right]$). We assumed that $\alpha$ is equal $0.1$, but changing of the value of $\alpha$ does not influence for results. Note that the values of the likelihood function are not sensitive to changing of $\frac{E_0}{3H_0^2}$.}
    \label{fig:fig14}
\end{figure}

\begin{figure}[ht]
    \centering
    \includegraphics[width=0.45\textwidth]{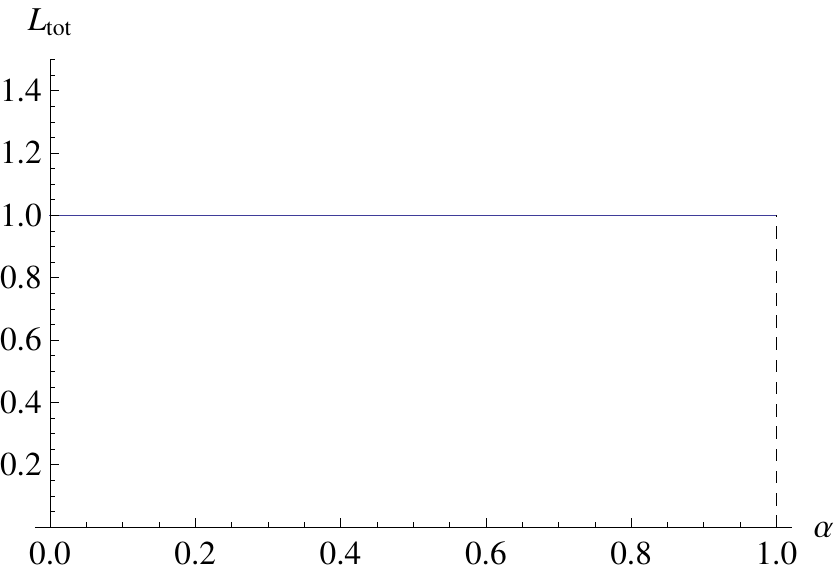}
    \caption{Diagram of PDF for parameter $\alpha$ obtained as an intersection of the likelihood function. Two planes of intersection likelihood function are $H_0=68.82 \left[\frac{\text{km}}{\text{s}\times\text{Mpc}}\right]$ and
$\Omega_{\text{m},0}=0.3009$. The planes of intersection are constructed from the best fitting value of the model parameters. We assume the value of $\alpha$ from the interval $(0,\text{ } 1)$. Note that the values of the likelihood function are not sensitive to changing of $\alpha$.}
    \label{fig:fig12}
\end{figure}

\begin{figure}[ht]
    \centering
    \includegraphics[width=0.45\textwidth]{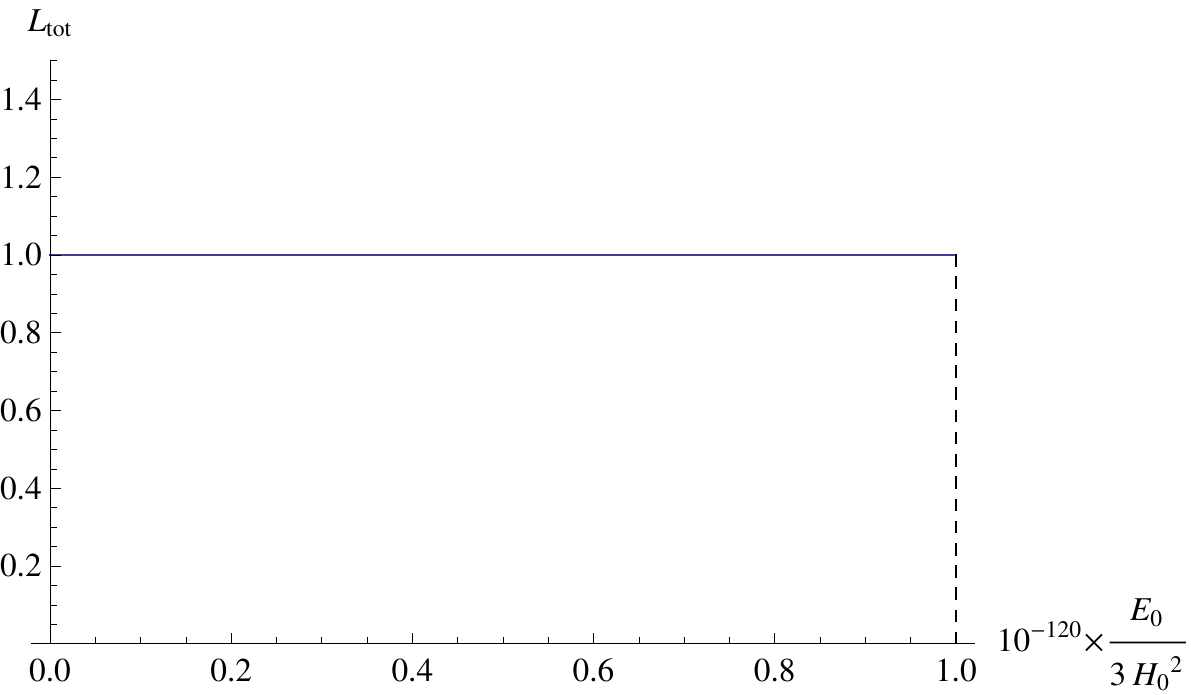}
    \caption{Diagram of PDF for $\frac{E_0}{3H_0^2}$ obtained as an intersection of the likelihood function. Two planes of intersection likelihood function are $H_0=68.82 \left[\frac{\text{km}}{\text{s}\times\text{Mpc}}\right]$ and $\Omega_{\text{m},0}=0.3009$. The planes of intersection are constructed from the best fitting value of the model parameters. We assume the value of $\frac{E_0}{3H_0^2}$ from the interval $(0,\text{ } 10^{120})$. Note that the values of the likelihood function are not sensitive to changing of $\frac{E_0}{3H_0^2}$.}
    \label{fig:fig13}
\end{figure}

\begin{table}
    \caption{The best fit and errors for the estimated model for SNIa+BAO+$H(z)$+AP+CMB test with $H_0$ from the interval (66.0, 72.0), $\Omega_{\text{m},0}$ from the interval $(0.27, 0.34)$. $\Omega_{\text{b},0}$ is assumed as 0.048468.}
    \label{table:1}
    \begin{center}
        \begin{tabular}{llll} \hline
            parameter & best fit & $ 68\% $ CL & $ 95\% $ CL  \\ \hline \hline
            $H_0$ & 68.82 km/(s Mpc) & $\begin{array}{c}
            +0.61 \\ -0.55
            \end{array}$ & $\begin{array}{c}
            +0.98 \\ -0.92
            \end{array}$ \\ \hline
            $\Omega_{\text{m},0}$ & 0.3009 & $\begin{array}{c}
            +0.0079 \\ -0.0084
            \end{array}$ & $\begin{array}{c}
            +0.0133 \\ -0.0134
            \end{array}$ \\ \hline
        \end{tabular}
    \end{center}
\end{table}

In this paper, the likelihood function for the measurements of CMB \cite{Ade:2015rim} and lensing by Planck, and low-$\ell$ polarization from the WMAP (WP) has the following form
\begin{equation}
\ln L_{\text{CMB}+\text{lensing}} = - \frac{1}{2}  (\mathbf{x}^{\text{th}}-\mathbf{x}^{\text{obs}})
\mathbb{C}^{-1} (\mathbf{x}^{\text{th}}-\mathbf{x}^{\text{obs}}),
\end{equation}
where $\mathbb{C}$ is the covariance matrix with the errors, $\mathbf{x}$ is a vector of the acoustic scale $l_{A}$, the shift parameter $R$ and $\Omega_{b}h^2$ where
\begin{align}
l_A &= \frac{\pi}{r_s(z^{*})} c \int_{0}^{z^{*}} \frac{dz'}{H(z')} \\
R &= \sqrt{\Omega_{\text{m,0}} H_0^2} \int_{0}^{z^{*}} \frac{dz'}{H(z')},
\end{align}
where $z^{*}$ is the redshift of the epoch of the recombination \cite{Hu:1995en}.

In this paper, the final formula for likelihood function is given in the following form
\begin{equation}
	L_{\text{tot}} = L_{\text{SNIa}} L_{\text{BAO}} L_{\text{AP}}
	L_{H(z)} L_{\text{CMB+lensing}}.
\end{equation}

The statistical analysis was done by our own code CosmoDarkBox. This code uses the Metropolis-Hastings algorithm \cite{Metropolis:1953am,Hastings:1970aa}.

We estimated four cosmological parameters: $H_0$, $\Omega_\text{m,0}$, $\alpha$ and $E_0$ parameter. Our statistical results are completed in Table~\ref{table:1}. We present intersections of the likelihood function with $68\%$ and
$95\%$ confidence level projections in Figs.~\ref{fig:fig6}-\ref{fig:fig14}. PDF diagrams for $\alpha$ and $\frac{E_0}{3H_0^2}$ are presented in Figs.~\ref{fig:fig12}-\ref{fig:fig13}.

The values of the likelihood function are not almost sensitive to changing of $\alpha$ and $E_0$ parameter. The possible changing of the values of the likelihood function are beyond abilities of numerical methods. This fact can be interpreted as the lack of sensitive of the present evolution of the universe for changing of $\alpha$ and $E_0$ parameter. The best fit values of $H_0$ and $\Omega_\text{m}$ for our model are equivalent of the best fit values for the $\Lambda$CDM model.

\section{Conclusion}

The main goal of our paper was to analyze the cosmological model with the running dark energy as well as the dark matter and the baryonic matter in the dust form. We considered the evolution of the dark energy using the fact that
the decay of  false vacuum to the true vacuum is the quantum decay process.
From the cosmological point of view this model was formulated in terms of the cosmological model with the interaction between the dark matter and the dark energy.

We detected the intermediate phase of oscillations between phases of the constant dark energy. The preceding phase has $\rho_{\text{de}}=E_0$ and the following phase has $\rho_\text{de}=\Lambda_\text{bare}$. Defining this class of models parametrized with $\alpha$ (the deviation from the $\Lambda$CDM model) we have found two different types of dynamical behaviour. Independly of $0<\alpha<1$ there is a universal mechanism of jumping of the value of energy density of dark energy from the initial value of $E_0 = 10^{120}$ to present value of the cosmological constant of $0.7$.

During this epoch there is the oscillatory behaviour of energy density of dark energy as well as its coefficient equation of state. In this intermediate regime the amplitude of oscillatory increases, then there is a jump down followed by the decreasing
oscillations. This kind of oscillation appears for $0<\alpha <0.4$. The number, period and amplitude of oscillations as well as the length of this intermediate regime decresess as the parameter $\alpha$ grows. For $\alpha > 0.4$ the oscilations disapear and only the jump down of energy density of dark energy remains. The jump down mechanism is independent from the parameter $\alpha$ value, which leads to to solving the cosmological constant problem.

In the early Universe the energy density of dark energy is significantly lower than the energy density of dark matter,
therefore
the change of energy density of the dark matter, which is caused by energy transfer in the dark sector, is negligible.

While our model make an attempt of an explanation of the cosmological constant problem, the coincidence problem is still open as we forced the model to have an exit on the present value of the cosmological constant.
In the early Universe, the dark energy oscillates. But the amplitude of oscillations decreases with time. In consequence for the late time Universe, oscillations are negligible and the dark energy can be described as the cosmological constant. Unfortunately our model cannot explain why the present value of dark energy and matter are comparable.

From the statistical analysis of the model we found that the model is generic in the sense that independently of the values of the parameters $\alpha$ and $E_0$ we can obtain the present value of the energy density of the dark energy. Therefore, the $\Lambda$CDM model is an attractor which the all models with different values of parameters $\alpha$ and $E_0$ can reach at. The final interval of evolution for which we have data at dispose is identical for whole class of models, therefore it is impossible to find best-fitted values of model parameters and indicate one model (degeneration problem).

As it should be expected it is difficult to discriminate the parameters of early state of the universe as there is no data for very high redshift. In Fig. \ref{fig:fig12} and \ref{fig:fig13} the likelihood functions for parameters of interest are flat, so there is no best fit value. That's why we take calibrated values of these parameters for further analysis in this paper. We assume that false vacuum energy is $10^{120}$ as is indicated from the quantum field theory. On other hand the parameter $\alpha$ should be chosen to get the decaying process of false vacuum to take place after the Planck era.

\begin{acknowledgements}
The work has been supported by Polish National Science Centre (NCN), project DEC-2013/09/B/ ST2/03455.
\end{acknowledgements}

\end{document}